\documentclass[preprint,showpacs,preprintnumbers,amsmath,amssymb,amssymb]{revtex4}

\usepackage{graphicx}
\usepackage{dcolumn}
\usepackage{bm}

\begin{document}

\hyphenpenalty=5000
\tolerance=1000
\makeatletter


\title{Charged Higgs in 3-3-1 Model Through $e^-e^+$ Collisions}

\author{J.\ E.\ Cieza Montalvo$^1$}
\affiliation{$^1$Instituto de F\'{\i}sica, Universidade do Estado do Rio de Janeiro, Rua S\~ao Francisco Xavier 524, 20559-900 Rio de Janeiro, RJ, Brazil}
\author{K. I. Cuba Quispe, R. J. Gil Ram\'{i}rez, C. A. Morgan Cruz, J. F. Rabanal Mu\~noz, G. H. Ram\'{i}rez Ulloa, A. I. Rivasplata Mendoza$^2$}
\affiliation{$^2$Universidad Nacional de Trujillo, Departamento de F\'{i}sica, Av. Juan Pablo II S/N; Ciudad Universitaria, Trujillo, La Libertad, Per\'u}

\author{M. D. Tonasse$^3$}
\affiliation{$^3$Instituto de F\'\i sica Te\'orica, Universidade Estadual Paulista, \\  Rua Dr. Bento Teobaldo Ferraz 271, 01140-070 S\~ao Paulo, SP, Brazil}
\date{\today}

\begin{abstract}
In this work we present an analysis of production and signature of charged Higgs bosons $H_2^{\pm}$ in the version of the 3-3-1 model containing heavy leptons at the CLIC (Cern Linear Collider). The production rate is found to be significant for the direct production of $e^{-} e^{+} \rightarrow  H_{2}^{+} H_{2}^{-}$. We also studied the possibility to identify it using their respective branching ratios.
\end{abstract}

\maketitle

\section{Introduction}

The Higgs sector still remains one of the most indefinite part of the standard model (SM) \cite{wsg}, but it still represents a fundamental rule by explaining how the particles gain masses by means of a isodoublet scalar field, which is responsible for the spontaneous breakdown of the gauge symmetry, the process by which the spectrum of all particles are generated. This process of mass generation is the so called {\it Higgs mechanism}, which plays a central role in gauge theories. \par

The SM provides a very good description of all the phenomena related to hadron and lepton colliders. This includes the Higgs boson which appears as elementary scalar and which arises through the breaking of electroweak symmetry. So, on $4$ July $2012$, the discovery of a new particle with a mass measured by CMS Collaboration to be $125.35$ GeV was announced; physicists suspected that it was the Higgs boson. Since then, the particle has been shown to behave, interact, and decay in many of the ways predicted for Higgs particles by the Standard Model, as well as having even parity and zero spin, two fundamental attributes of a Higgs boson. This also means it is the first elementary scalar particle discovered in nature, \cite{cms1}. Any discovery of a charged Higgs boson would be confirmation of new physics and the CLIC can admit or exclude such a probability, as long as the Higgs had the energy reached by this accelerator. \par

Different types of Higgs bosons, if they exist, may lead us into new realms of physics beyond the SM. Since the SM leaves many questions open, there are several  extensions. For example, if the Grand Unified Theory (GUT) contains the SM at high energies, then the Higgs bosons associated with GUT symmetry breaking must have masses of order $M_{X} \sim {\cal O} (10^{15})$ GeV. Supersymmetry \cite{supers} provides a solution to hierarchy problem through the cancellation of the quadratic divergences via fermionic and bosonic loops contributions \cite{cancell}. Moreover, the Minimal Supersymmetric extension of the SM can be derived as an effective theory of supersymmetric GUT \cite{sgut}. \par

There are also other class of models based on SU(3)$_C \otimes$SU(3)$_L \otimes$U(1)$_N$ gauge symmetry (3-3-1 model) \cite{PT93a, PP92, FR92}, where the anomaly cancellation mechanisms occur when the three basic fermion families are considered and not family by family as in the SM. This mechanism is peculiar because it requires that the number of families is an integer multiple of the number of colors. This feature combined together  with the asymptotic freedom, which is a property of quantum chromodynamics, requires that the number of families is three. Moreover, according to these models, the Weinberg angle is restricted to the value $s_W^2 = \sin^2\theta_W <1/4$ in the version of heavy-leptons \cite {PT93a}. Thus, when it evolves to higher values, it shows that the model loses its perturbative character when it reaches to mass scale of about 4 TeV \cite{DI05}. Hence, the 3-3-1 model is one of the most interesting extensions of the SM and is phenomenologically well motivated to be probed at the CLIC and other accelerators. \par

Among the most studied candidate theories for electroweak symmetry breaking (EWSB) in the literature are the Higgs mechanism within the Standard Model (SM) \cite{glashow}, the Minimal Supersymmetric Standard Model (MSSM), \cite{nilles} and the 3-3-1 Model \cite{TO96}. Contrary to the SM and MSSM, three Higgs scalar triplets are required in the 3-3-1 Model. This results in seven physical Higgs bosons instead of the single Higgs boson in the SM and five in the MSSM,  these states are: three scalar ($H_0^1$, $H_0^2$, $h_0$), one pseudoscalar neutral ($H_0^3$) and three charged Higgs bosons ($H_1^{\pm}$, $H_2^{\pm}$, $H^{\pm \pm}$). \par

The outline of this work is the following. In Sec. II we present the relevant features of the model. In Sec. III we compute the total cross sections of the process and the respective branching ratios of $H_2^{\pm}$ and in Sec. IV we summarize our results and conclusions. \par


\section{Relevant Features of the Model \label{sec2}}

If there exist any model beyond the standard model (BSM), then it is no strange that it is still hidden in the
scalar sector. As one of the simplest BSM models, is the $3-3-1$ model, which has been extensively investigated in the literature. \par

These is a promissory model which is based on the $SU(3)_{C}\otimes  SU(3)_{L} \otimes U(1)_{N}$ (3-3-1 for short) semi simple symmetry group \cite{PT93}, which contains both charged Higgs bosons. In this model the three Higgs scalar triplets
\begin{equation}
\eta = \left(\begin{array}{c} \eta^0 \\  \eta_1^- \\  \eta_2^+ \end{array}\right) \sim \left({\bf 3}, 0\right), \quad \rho = \left(\begin{array}{c} \rho^+ \\  \rho^0 \\  \rho^{++}
\end{array}\right) \sim \left({\bf 3}, 1\right), \quad \chi =
\left(\begin{array}{c} \chi^- \\
\chi^{--} \\ \chi^0 \end{array}\right) \sim \left({\bf 3}, -1\right),
\label{higgs}\end{equation}
generate the fermion and gauge boson masses in the model. The neutral scalar fields develop the vacuum expectation values (VEVs)  $\langle\eta^0\rangle = v_\eta$, $\langle\rho^0\rangle = v_\rho$ and  $\langle\chi^0\rangle = v_\chi$, with $v_\eta^2 + v_\rho^2 = v_W^2 = (246 \mbox{ GeV})^2$. The pattern of symmetry breaking is
\[ \mbox{SU(3)}_L \otimes\mbox{U(1)}_N \stackrel{\langle\chi\rangle}{\longmapsto}\mbox{SU(2)}_L\otimes\mbox{U(1)}_Y
\stackrel{\langle\eta, \rho\rangle}{\longmapsto}\mbox{U(1)}_{\rm em}\]
and so, we can expect $v_\chi \gg v_\eta, v_\rho$. The $\eta$ and $\rho$ scalar triplets give masses to the ordinary fermions and gauge bosons, while the $\chi$ scalar triplet gives masses to the new fermions and new gauge bosons. The most general, gauge invariant and renormalizable Higgs potential is
\begin{eqnarray}
V\left(\eta, \rho, \chi\right) & = & \mu_1^2\eta^\dagger\eta + \mu_2^2\rho^\dagger\rho + \mu_3^2\chi^\dagger\chi + \lambda_1\left(\eta^\dagger\eta\right)^2 + \lambda_2\left(\rho^\dagger\rho\right)^2 + \lambda_3\left(\chi^\dagger\chi\right)^2 + \cr
&& \left(\eta^\dagger\eta\right)\left[\lambda_4\left(\rho^\dagger\rho\right) + \lambda_5\left(\chi^\dagger\chi\right)\right] + \lambda_6\left(\rho^\dagger\rho\right)\left(\chi^\dagger\chi\right) + \lambda_7\left(\rho^\dagger\eta\right)\left(\eta^\dagger\rho\right) + \cr
&& \lambda_8\left(\chi^\dagger\eta\right)\left(\eta^\dagger\chi\right) + \lambda_9\left(\rho^\dagger\chi\right)\left(\chi^\dagger\rho\right) + \lambda_{10}\left(\eta^\dagger\rho\right)\left(\eta^\dagger\chi\right) + \cr
&& \frac{1}{2}\left(f\epsilon^{ijk}\eta_i\rho_j\chi_k + {\mbox{H. c.}}\right).
\label{pot}\end{eqnarray}
Here $f$ is a constant with dimensions of mass and the $\lambda_i$, $\left(i = 1, \dots, 10\right)$ are adimensional constants with $\lambda_3 < 0$ from the positivity of the scalar masses. The term proportional to $\lambda_{10}$ violates lepto-barionic number so that, it was not considered in the analysis of the Ref. \cite{TO96} (another analysis of the 3-3-1 scalar sector are given in Ref. \cite{AK00} and references cited therein). We can notice that this term contributes to the mass matrices of the charged scalar fields, but not to the neutral ones.  However, can be checked that in the approximation $v_\chi \gg v_\eta, v_\rho$ we can still work with the masses and eigenstates given in Ref. \cite{TO96}. Here this term is important to the decay of the lightest exotic fermion. Therefore, we are keeping it in the Higgs potential.\par
Symmetry breaking is initiated when the scalar neutral fields are shifted as $\varphi = v_\varphi + \xi_\varphi + i\zeta_\varphi$, with $\varphi$ $=$  $\eta^0$, $\rho^0$, $\chi^0$. Thus, the physical neutral scalar eigenstates  $H^0_1$, $H^0_2$, $H^0_3$ and $h^0$ are related to the shifted fields as
\begin{mathletters}\begin{equation}
\left(\begin{array}{c} \xi_\eta \\  \xi_\rho \end{array}\right) \approx
\frac{1}{v_W}\left(\begin{array}{cc} v_\eta & v_\rho \\  v_\rho & -v_\eta
\end{array}\right)\left(\begin{array}{c} H^0_1 \\  H^0_2 \end{array}\right), \qquad \xi_\chi \approx H^0_3, \qquad \zeta_\chi  \approx h^0,
\label{eign}\end{equation}
and in the charge scalar sector we have

\begin{equation}
\eta^+_1 \approx \frac{v_\rho}{v_W}H^+_1, \qquad \rho^+ \approx \frac{v_\eta}{v_W}H_2^+, \qquad \chi^{++} \approx \frac{v_\rho}{v_\chi}H^{++},
\label{eigc}\end{equation}\label{eig}\end{mathletters}
with the condition that $v_\chi \gg v_\eta, v_\rho$ \cite{TO96}.

In this work, we will study the production mechanism for Higgs particles ($H^{-} H^{+}$) and their signatures in $e^{+} e^{-}$ colliders such as the CERN Linear Collider (CLIC) ($\sqrt{s} = 3.0$ and $5.0$ ) \ TeV. \par


\section{Cross section production}
\label{secIII}

The process $e^{-} e^{+} \rightarrow  H^{-}_{2} H^{+}_{2}$, take place through the exchange of the bosons $\gamma$, $Z$ and $Z^\prime$ in the $s$ channel. The model does not admit interaction between the neutral and charged Higgs. Then using the interaction Lagrangians written above we evaluate the differential cross section for the $H_{2}^{\pm}$

\begin{equation}
\frac{d \hat{\sigma}}{d \Omega} = \frac{1}{64 \pi^{2} \hat{s}} \left(\overline{\left|M_{\gamma} \right|^2} + \overline{\left|M_{Z}\right|^{2}} + \overline{\left|M_{Z'} \right|^{2}} + 2 {\it Re} \overline{M_{\gamma}M_Z} +  2 {\it Re} \overline{M_{\gamma}M_Z'} + 2 {\it Re} \overline{M_{Z}M_Z'}       \right).
\end{equation}

Writing separately the cross-sections, we will have

\begin{eqnarray}
\frac{d \hat{\sigma}_{\gamma}}{d\cos \theta} & = &\frac{\beta_{H_2^{\pm}} \alpha^{2} \pi (\Lambda_{\gamma H_2^- H_2^+})^{2}}{16 s^{3} \sin^{2}_{\theta_{W}}} \left [(s- 4m_{H_{2}^{\pm}}^{2})s- (t- u)^2 \right ]
\end{eqnarray}

\begin{eqnarray}
\frac{d \hat{\sigma}_{Z}}{d\cos \theta} & = &\frac{\beta_{H_2^{\pm}} \alpha^{2} \pi (\Lambda_{Z H_2^- H_2^+})^{2} (g_{V}^{2} + g_{A}^{2})} {64 \sin^{4}_{\theta_{W}} \cos^{2}_{\theta_{W}} s (s-m_{Z}^{2}+ im_{Z} \Gamma_{Z})^{2}}   \nonumber  \\
&& \hskip 5.0cm  \left [(s- 4m_{H_{2}^{\pm}}^{2})s- (t- u)^2 \right ]
\end{eqnarray}

\begin{eqnarray}
\frac{d \hat{\sigma}_{Z^{\prime}}}{d\cos \theta} & = &\frac{\beta_{H_2^{\pm}} \alpha^{2} \pi (\Lambda_{Z^\prime H_2^- H_2^+})^{2} (g_{V'}^{2} + g_{A'}^{2})} {192 \sin^{4}_{\theta_{W}} \cos^{2}_{\theta_{W}} s (s-m_{Z'}^{2}+ im_{Z'} \Gamma_{Z'})^{2}}   \nonumber  \\
&& \hskip 5.0cm  \left [(s- 4m_{H_{2}^{\pm}}^{2})s- (t- u)^2 \right ]
\end{eqnarray}

\begin{eqnarray}
\frac{d \hat{\sigma}_{\gamma - Z}}{d\cos \theta} & = & -\frac{\beta_{H_2^{\pm}} \alpha^{2} \pi (\Lambda_{\gamma H_2^- H_2^+})(\Lambda_{Z H_2^- H_2^+}) (g_{V})} {16 \sin^{3}_{\theta_{W}} \cos_{\theta_{W}} s^2 (s-m_{Z}^{2}+ im_{Z} \Gamma_{Z})}   \nonumber  \\
&& \hskip 5.0cm   \left [(s- 4m_{H_{2}^{\pm}}^{2})s- (t- u)^2 \right ]
\end{eqnarray}

\begin{eqnarray}
\frac{d \hat{\sigma}_{\gamma - Z^{\prime}}}{d\cos \theta} & = & -\frac{\beta_{H_2^{\pm}} \alpha^{2} \pi (\Lambda_{\gamma H_2^- H_2^+})(\Lambda_{Z^{\prime} H_2^- H_2^+}) (g_{V^{\prime}})} {16 \sqrt{3} \sin^{3}_{\theta_{W}} \cos_{\theta_{W}} s^2 (s-m_{Z^{\prime}}^{2}+ im_{Z^{\prime}} \Gamma_{Z^{\prime}})}   \nonumber  \\
&& \hskip 5.0cm   \left [(s- 4m_{H_{2}^{\pm}}^{2})s- (t- u)^2 \right ]
\end{eqnarray}

\begin{eqnarray}
\frac{d \hat{\sigma}_{Z - Z^{\prime}}}{d\cos \theta} & = & -\frac{\beta_{H_2^{\pm}} \alpha^{2} \pi (\Lambda_{Z H_2^- H_2^+})(\Lambda_{Z^{\prime} H_2^- H_2^+})} {32 \sqrt{3} \sin^{4}_{\theta_{W}} \cos^{2}_{\theta_{W}} s (s-m_{Z}^{2}+ im_{Z} \Gamma_{Z}) (s-m_{Z^{\prime}}^{2}+ im_{Z^{\prime}} \Gamma_{Z^{\prime}})}  \nonumber  \\
&&  \hskip 3.5cm (g_{V} g_{V'} + g_{A} g_{A'})  \left [(s- 4m_{H_{2}^{\pm}}^{2})s- (t- u)^2 \right ]
\end{eqnarray}

the $\beta_{H_{2}^{\pm}}$ is the Higgs velocity in the c.m. of the process which is equal to
\[
\beta_{H_{2}^{\pm}} = \sqrt{1- \frac{4m_{H_{2}^{\pm}}^2}{s}}  \ \ ,
\]

and $t$ and $u$ are

\[
t  = m_{H_{2}^{\pm}}^{2} - \frac{s}{2}  \left(1- \beta_{H_2^{\pm}} \cos \theta  \right)  ,
\]

\[
u  = m_{H_{2}^{\pm}}^{2} - \frac{s}{2}  \left(1+ \beta_{H_2^{\pm}} \cos \theta  \right)  ,
\]
where $\theta$ is the angle between the Higgs and the incident electron in the CM frame. \par

The primes $\left(^\prime\right)$ are for the case when we take a $Z'$ boson, $\Gamma_{Z'}$ \cite{ct2005, cieto02}, are the total width of the $Z'$ boson, $g_{V, A}$ are the standard lepton coupling constants, $g_{V', A'}$ are the 3-3-1 lepton coupling constants, $s$ is the center of mass energy of the $e^{-} e^{+}$ system, $g= \sqrt{4 \ \pi \ \alpha}/\sin \theta_{W}$ and $\alpha$ is the fine structure constant, which we take equal to $\alpha=1/128$. For the $Z^\prime$ boson we take  $m_{Z^\prime} = \left(5.9 - 8.6\right)$ TeV \cite{bhupal, okada} , since $m_{Z^{\prime}}$ is proportional to the VEV $v_\chi$ \cite{TO96,PP92,FR92}. For the standard model parameters, we assume Particle Data Group values, {\it i. e.}, $m_Z = 91.19$ GeV, $\sin^2{\theta_W} = 0.2315$, and $M_W = 80.33$ GeV  \cite{Nea10}, the velocity of the Higgs in the c.m of the process we denote through $\beta_{H_{2}^{\pm}}$, $\it{t}$ and $\it{u}$ are the kinematic invariants. We have also defined the $\Lambda_{\gamma H_2^- H_2^+}$ as the coupling constants of the $\gamma$ to Higgs $H_{2}^{\pm}$, the $\Lambda_{Z H_2^- H_2^+} $ as the coupling constants of the $Z$ to Higgs $H_{2}^{\pm}$ and the $\Lambda_{Z^{\prime} H_2^- H_2^+}$ as being the coupling constants of the $Z^{\prime}$ to Higgs $H_{2}^{\pm}$. \par

\[
\Lambda_{\gamma H_2^- H_2^+} = \frac{t_{W} (v_{\chi}^{2}- v_{\rho}^{2})}{ \sqrt{1+4t_{W}^{2}} \ (v_{\chi}^{2}+ v_{\rho}^{2})}
\]

\[
\Lambda_{Z H_2^- H_2^+} = -\frac{v_{\rho}^{2} (1+ 6t_W^2)- 2t^2 v_{\chi}^2}{
(v_{\chi}^2 + v_{\rho}^2) \sqrt{(1+4t_{W}^{2})(1+ 3t_{W}^{2})}}
\]

\[
\Lambda_{Z^{\prime} H_2^- H_2^+} = \frac{v_{\rho}^{2} (6t_W^2-1)+ 2 v_{\chi}^2}{(v_{\chi}^2 + v_{\rho}^2) \sqrt{(1+ 3t_{W}^{2})}}
\]

The total width of the Higgs $H_{2}^{-}$ into quarks, leptons, gauge bosons $V^{-} \gamma$, $Z V^{-}$, $Z^{\prime} V^{-}$, gauge bosons and Higgs bosons $W^{+} H^{--}$, $V^{-} H_1^0$, $V^{-} H_2^0$, $V^{-} H_3^0$, $V^{-} h^0$, are,
respectively, given by
\begin{eqnarray}
\Gamma \left(H_{2}^{\pm} \to {\rm all}\right)&  = & \Gamma_{H_{2}^{\pm} \to q \bar{q}}  +   \Gamma_{H_{2}^{\pm} \to E_a^{\pm} \nu(\bar{\nu})} +  \Gamma_{H_{2}^{\pm} \to V^{\pm} \gamma} + \Gamma_{H_{2}^{\pm} \to Z V^{\pm}} + \Gamma_{H_{2}^{\pm} \to Z^{\prime} V^{\pm}} +  \Gamma_{H_{2}^{\pm} \to W^{\mp} H^{\pm \pm} }  \nonumber  \\
&&+ \Gamma_{H_{2}^{\pm} \to H^{0}_{1} V^{\pm}} + \Gamma_{H_{2}^{\pm} \to  H^{0}_{2} V^{\pm}} + \Gamma_{H_{2}^{\pm} \to  H^{0}_{3} V^{\pm}} + \Gamma_{H_{2}^{\pm} \to  h^{0} V^{\pm}}    ,
\end{eqnarray}
where we have for each the widths given above that:

\begin{mathletters}
\begin{eqnarray}
\Gamma_{H_{2}^{\pm} \to q \bar{q}} & = & \frac{3 \sqrt{1- \left (\frac{m_{J_1} + m_U}{m_{H_2^{\pm}}} \right )^2} \sqrt{1- \left (\frac{m_{J_1} - m_U}{m_{H_2^{\pm}}} \right )^2} m_U^2 v_{\chi}^2 }{16 \pi m_{H_{2}^{\pm}} v_{\rho}^2 (v_{\rho}^2 + v_{\chi}^2) }  (s-m_{J_1}^2 -  m_{U}^{2} ) ,  \\
\Gamma_{H_{2}^{\pm} \to E_a^{\pm} \nu(\bar{\nu})} & = &\frac{\sqrt{1- \left (\frac{m_{E^{\pm}} + m_{\nu}}{m_{H_2^{\pm}}} \right )^2} \sqrt{1- \left (\frac{m_{E^{\pm}} - m_{\nu}}{m_{H_2^{\pm}}} \right )^2} m_{E^{\pm}}^2 v_{\rho}^2}{16 \pi m_{H_{2}^{\pm}} v_{\chi}^2 (v_{\rho}^2 + v_{\chi}^2) }  (s-m_{E^{\pm}}^2 -  m_{\nu}^{2} ) ,  \\
\Gamma_{H_{2}^{\pm} \to V{\pm} \gamma} & = & \frac{3 \sqrt{1-  m^2_{V^{\pm}}/  m^{2}_{H_{2}^{\pm}}} e^4 v_{\rho}^2 v_{\chi}^2}{32 \pi m_{H_{2}^{\pm}}  (v_{\rho}^2 + v_{\chi}^2) }  ,  \\
\Gamma_{H_{2}^{\pm} \to Z V^{\pm}} & = & \frac{\sqrt{1- \left (\frac{m_{Z} + m_V}{m_{H_2^{\pm}}} \right )^2} \sqrt{1- \left (\frac{m_{Z} - m_V}{m_{H_2^{\pm}}} \right )^2}}{16 \pi \cos^{2}_{\theta_{W}} \sin^{4}_{\theta_{W}}  m_{H_{2}^{\pm}}} \frac{e^4 (1+ \sin^{2}_{\theta_{W}})^2 v_{\rho}^2 v_{\chi}^2}{2(v_{\rho}^2 v_{\chi}^2) }
\left (\frac{5}{2}+ \frac{1}{4} \frac{m_{Z}^{4}}{m_{V}^{2}} + \frac{1}{4} \frac{m_{V}^{2}}{m_{Z}^{2}} \right.\nonumber  \\
&& \left.+ \frac{1}{4} \frac{m_{{H}_{2}^{\pm}}^{4}}{m_{Z}^{2} m_{V}^{2}} - \frac{1}{2} \frac{m_{{H}_{2}^{\pm}}^{2}}{m_{Z}^{2}} - \frac{1}{2} \frac{m_{{H}_{2}^{\pm}}^{2}}{m_{V}^{2}} \right ) ,   \\
\Gamma_{H_{2}^{\pm} \to W^{\mp} H^{\pm \pm} } & = & \frac{\sqrt{1- \left (\frac{m_{W^{\mp}} + m_{H^{\pm \pm}}}{m_{H_2^{\pm}}} \right )^2} \sqrt{1- \left (\frac{m_{W^{\mp}} - m_{H^{\pm \pm}}}{m_{H_2^{\pm}}} \right )^2} e^2 v_{\rho}^2 v_{\eta}^2}{ 128 \pi \sin^{2}_{\theta_{W}} m_{H_{2}^{\pm}} (v_{\rho}^2 + v_{\chi}^2) (v_{\eta}^2 + v_{\chi}^2) }
\left (\frac{m_{W}^{2}}{4} + \frac{m_{H_{2}^{\pm}}^{4}}{4m_{W}^{2}}- \frac{m_{H_{2}^{\pm}}^{2} m_{H^{\pm \pm}}^{2}}{2m_{W}^{2}}   \right.\nonumber  \\
&&\left.+\frac{ m_{H^{\pm \pm}}^{4}}{4m_{W}^{2}} - \frac{m_{H_{2}^{\pm}}^{2}}{2} - \frac{m_{H^{\pm \pm}}^{2}}{2} \right ) ,  \\
\Gamma_{H_{2}^{\pm} \to V^{\pm} H_{1}^{0} } & = & \frac{\sqrt{1- \left (\frac{m_{V^{\pm}} + m_{H_1^{0}}}{m_{H_2^{\pm}}} \right )^2} \sqrt{1- \left (\frac{m_{V^{\pm}} - m_{H_1^{0}}}{m_{H_2^{\pm}}} \right )^2} e^2 v_{\rho}^2 v_{\chi}^2}{ 128 \pi \sin^{2}_{\theta_{W}} m_{H_{2}^{\pm}} v_{\chi}^2 (v_{\rho}^2 + v_{\chi}^2) }
\left (\frac{m_{V}^{2}}{4} + \frac{m_{H_{2}^{\pm}}^{4}}{4m_{V}^{2}}- \frac{m_{H_{2}^{\pm}}^{2} m_{H_{1}^{0}}^{2}}{2m_{V}^{2}}   \right.\nonumber  \\
&&\left.+\frac{ m_{H_{1}^{0}}^{4}}{4m_{V}^{2}} - \frac{m_{H_{2}^{\pm}}^{2}}{2} - \frac{m_{H_{1}^{0}}^{2}}{2} \right ) ,  \\
\Gamma_{H_{2}^{\pm} \to V^{\pm} H_{2}^{0} } & = & \frac{\sqrt{1- \left (\frac{m_{V^{\pm}} + m_{H_2^{0}}}{m_{H_2^{\pm}}} \right )^2} \sqrt{1- \left (\frac{m_{V^{\pm}} - m_{H_2^{0}}}{m_{H_2^{\pm}}} \right )^2} e^2 v_{\eta}^2 v_{\chi}^2}{ 128 \pi \sin^{2}_{\theta_{W}} m_{H_{2}^{\pm}} v_{\chi}^2 (v_{\rho}^2 + v_{\chi}^2) }
\left (\frac{m_{V}^{2}}{4} + \frac{m_{H_{2}^{\pm}}^{4}}{4m_{V}^{2}}- \frac{m_{H_{2}^{\pm}}^{2} m_{H_{2}^{0}}^{2}}{2m_{V}^{2}}   \right.\nonumber  \\
&&\left.+\frac{ m_{h^{0}}^{4}}{4m_{V}^{2}} - \frac{m_{H_{2}^{\pm}}^{2}}{2} - \frac{m_{h^{0}}^{2}}{2} \right ) ,  \\
\Gamma_{H_{2}^{\pm} \to V^{\pm} H_{3}^{0} } & = & \frac{\sqrt{1- \left (\frac{m_{V^{\pm}} + m_{H_3^{0}}}{m_{H_2^{\pm}}} \right )^2} \sqrt{1- \left (\frac{m_{V^{\pm}} - m_{H_3^{0}}}{m_{H_2^{\pm}}} \right )^2} e^2 v_{\rho}^2}{ 128 \pi \sin^{2}_{\theta_{W}} m_{H_{2}^{\pm}} (v_{\rho}^2 + v_{\chi}^2) }
\left (\frac{m_{V}^{2}}{4} + \frac{m_{H_{2}^{\pm}}^{4}}{4m_{V}^{2}}- \frac{m_{H_{2}^{\pm}}^{2} m_{H_{3}^{0}}^{2}}{2m_{V}^{2}}   \right.\nonumber  \\
&&\left.+\frac{ m_{H_{3}^{0}}^{4}}{4m_{V}^{2}} - \frac{m_{H_{2}^{\pm}}^{2}}{2} - \frac{m_{H_{3}^{0}}^{2}}{2} \right ) \nonumber \\  ,
\end{eqnarray}
\end{mathletters} \noindent

\begin{eqnarray}
\Gamma_{H_{2}^{\pm} \to V^{\pm} h^{0} } & = & \frac{\sqrt{1- \left (\frac{m_{V^{\pm}} + m_{h^{0}}}{m_{H_2^{\pm}}} \right )^2} \sqrt{1- \left (\frac{m_{V^{\pm}} - m_{h^{0}}}{m_{H_2^{\pm}}} \right )^2} e^2 v_{\eta}^2 v_{\chi}^2}{ 128 \pi \sin^{2}_{\theta_{W}} m_{H_{2}^{\pm}} v_{\chi}^2 (v_{\rho}^2 + v_{\chi}^2) }
\left (\frac{m_{V}^{2}}{4} + \frac{m_{H_{2}^{\pm}}^{4}}{4m_{V}^{2}}- \frac{m_{H_{2}^{\pm}}^{2} m_{h^{0}}^{2}}{2m_{V}^{2}}   \right.\nonumber  \\
&&\left.+\frac{ m_{h^{0}}^{4}}{4m_{V}^{2}} - \frac{m_{H_{2}^{\pm}}^{2}}{2} - \frac{m_{h^{0}}^{2}}{2}  \right ) \nonumber \\
\end{eqnarray}


\section{Results and conclusions}

Here we present the cross section for the process $e^{-} e^{+} \rightarrow  H^{-}_{2} H^{+}_{2}$ for  the CLIC ($3$ and $5$ TeV). All calculations were done according to \cite{tona96} from which we obtain for the parameters and the VEV, the following representative values:  $\lambda_{1} = 1.54 \times 10^{-1} $,  $\lambda_{2}=1.0$, $\lambda_{3}= -2.5 \times 10^{-2} $, $\lambda_{4}=  2.14$, $\lambda_{5}=-1.57$, $\lambda_{6}= 1.0$, $\lambda_{7} =-2.0$, $\lambda_{8}=-5.0 \times 10^{-1} $,  $v_{\eta}=195$ GeV, and $\lambda_{9}=0.0(0.0, 0.0)$ which correspond to $v_\chi= 3.5(4.0, 5.0)$ TeV, these parameters and VEV are used to estimate the values for the particle masses which are given in table \ref{tab1}. In particular the value of $v_{\eta}=195$ GeV, was taken to fix the mass of $H_ 1^0$.
\par

Differently from what we did in other's papers \cite{ct2005, cieto03, cieto02}, where was taken arbitrary parameters, in this work we take for the parameters and the VEV the following representative values given above and also the values of the mass of $H_1^0$ which is already defined \cite{atlas1, atlas11}. It is remarkable that the cross sections were calculated in order to guarantee the approximation $-f \simeq v_\chi$ \cite{tona96}.  It must be taken into consideration that the branching ratios of $H^{\pm}_{2}$ are dependent on the parameters of the 3-3-1, which determines the size of several decay modes. \par
	
\begin{table}[h]
\caption{\label{tab1} Values for the particle masses used in this work.
All the values in this Table are given in GeV. Here, $m_{H^{\pm\pm}} =
2825.0 (3227.7; 4035.5)$ GeV for $v_{\chi}=3500.0(4000.0, 5001.0)$ GeV and $m_T = 2v_\chi$.}
\begin{ruledtabular}
\begin{tabular}{c|cccccccccccccc}
$f$ & $v_{\chi}$, $m_{J_1}$ & $m_E$ & $m_M$  & $m_{H_3^0}$ & $m_{h^0}$ &
$m_{H_1^0}$ & $m_{H_2^0}$  & $m_V$ & $m_U$ & $m_{Z^\prime}$
& $m_{J_{2, 3}}$ \\
\hline
-3500 & 3500 & 521.15 & 3062.60  & 1106.80 & 5037.72 & 125.51 & 3560.15   &
  1608.59 & 1607.57 & 5976.43 & 4949.70 \\
-4000 & 4000 & 595.60 & 3500.02  & 1264.91 & 5756.99 & 125.51 & 4068.75   &
  1837.72 & 1836.83 & 6830.21 & 5656.80 \\
-5000 & 5000 & 744.50 & 4375.00  & 1581.46 & 7198.47 & 125.51 & 5086.95   &
  2296.62 & 2295.91 & 8539.47 & 7071.00 \\
\end{tabular}
\end{ruledtabular}
\end{table}

The Higgs $H^{\pm}_{2}$ in 3-3-1 model is coupled to quarks, $q \bar{q}$; heavy leptons and neutrinos, $E_{i}^{\pm} \nu(\bar{\nu})$; charged bosons and photons, $V^{\pm} \gamma$; gauge bosons and charged bosons, $Z V^{\pm}$; charged gauge bosons and double charged Higgs bosons, $W^{\pm} H^{\mp \mp}$; charged bosons and neutral Higgs, $V^{\pm} H_{i}^{0}$ where $i=1,2,3$. The Higgs $H^{\pm}_{2}$ can be heavier than $1190.65$ GeV for $v_{\chi} = 3.5$ TeV; $1360.51$ GeV for $v_{\chi} = 4.0$ teV and $1698.72$ GeV for $v_{\chi} = 5.0$ TeV, so the Higgs $H^{\pm}_{2}$ is a heavy particle for the $v_{\chi}$ taken above. \par

Previous ATLAS results exclude a $Z^{\prime}$ with mass less than $2.20$ TeV at 95  $\%$  C.L. Table I show the masses of the exotic boson $Z^{\prime}$, taken above, which are in accord with the estimates of the ATLAS \cite{atlas2014}.

\subsection{CLIC with 3 TeV }

We will start the phenomenology for the click, because the energy of the vacuum expectation value $v_{\chi}$ and therefore the masses, that it generates are not supported by the ILC (International Linear Collider). \par

Considering that the expected integrated luminosity for CLIC collider will be of order of $5$ ab$^{-1}$/yr, then we obtain a total of $ \simeq 2.04 \times 10^4 (0.00; 0.00)$ events per year if we take the mass of the Higgs boson $m_{H_2^{\pm}}= 1.3$ TeV and  $v_{\chi}=3.5(4.0; 5.0)$ TeV, and $1.31 \times 10^3 (8.35 \times 10^2; 0.00)$ events per year if we take the mass of the Higgs boson $m_{H_2^{\pm}}= 1.47$ TeV, please see Fig. 1, for the same vacuum expectation values $v_{\chi}$, taken above. The value of $0.0$ is for $v_{\chi}=4.0(5.0)$ TeV and mass $m_{H_2^{\pm}}= 1.3$ TeV, this values of $v_{\chi}$ restricts Higgs mass to values up to $1360.51(1698.72)$ GeV. The total widths for these vacuum expectation values are, $\Gamma_{H^{\pm}_{2}} = 8.08 \times 10^{-4} (0.00; 0.00)$ for $m_{H_2^{\pm}}= 1.3$ TeV and $\Gamma_{H^{\pm}_{2}} = 9.71 \times 10^{-4} (6.96 \times 10^{-4}; 0.00)$ for $m_{H_2^{\pm}}= 1.47$ TeV, see Fig. 2. \par

The impact of the initial state radiation (ISR) and beamstrahlung (BS) on precision measurements strongly affects the behaviour of the production cross section around the resonance peaks, modifying as the shape as the size \cite{kuraev}, so Fig.$3$, $4$ and $5$ shows the cross section with and without ISR + BS around the resonances points:  $m_{Z^\prime} = 5976.43$ GeV, $m_{Z^\prime} = 6830.21$ GeV and $m_{Z^\prime} = 8539.47$ GeV for $v_{\chi}=3.5(4.0; 5.0)$ TeV. As can be seen the peak of the resonance shifts to the right and is lowered as a  result of the ISR + BS effects, \cite{morgan2}.   \par

To obtain event rates, we multiply the production cross sections by the respective branching ratios. The mains decays of $H_2^{\pm}$, which are allowed for $v_{\chi}=3.5$ TeV and $m_{H^{\pm}}= 1.3$ TeV are $E^{\pm} \nu(\bar{\nu})$ (100.00 \ \%) whose number of events are $\simeq 2.04 \times 10^4$ and for $m_{H^{\pm}}= 1.47$ TeV the number of events will be $\simeq 1.31 \times 10^3$, and for $v_{\chi}=4.0$ TeV and $m_{H^{\pm}}= 1.47$ TeV, the number of events will be $\simeq 8.35 \times 10^2$, respectively, recall that for $v_{\chi}=4.0$ the appropriate masses begin with $m_{H^{\pm}}= 1.36$ TeV, see Fig. $6$ and Fig. $7$. \par

These two charged leptons $E^{\mp} E^{\pm}$ are relatively easy to measure in the detector, as they leave tracks and do not shower like gluons and quarks, by computing the invariant masses of both the pairs where a sharp charged Higgs bosons will be observed in the two opposite-sign lepton invariant mass distribution. Furthermore, the backgrounds events also can be readily reduced if we impose the Z window cut where the invariant mass of opposite-sign lepton pairs must be far from the Z mass:\ $|m_{E^\mp E^\pm}-m_{Z}|> 10$ GeV, this removes events where the leptons come from the Z decay \cite{cheng}. Whit respect to the $\bar{\nu} (\nu)$ we put the cut on the missing transverse momentum ${p\!\!\slash}_{T} >$ 20 GeV, both allows for a very strong reduction of the backgrounds. \par

 The processes that have the same signature of the signal $E^{+} E^{-} \nu \bar{\nu}$ are $e^{+} e^{-} \rightarrow \tau^{+} \tau^{-}, W^{+} W^{-}, Z^{0} Z^{0}$ and $W^{+} W^{-} Z$, the last process  is suppressed at least by $\alpha/ sin^{2} \theta_W$ relative to the process involving a double gauge boson, so using the COMPHEP \cite{pukhov}, the total cross section for these processes for $3.0$ TeV will be equal to $5.33 \times 10^{-1}$ pb, consequently the number of backgrounds events will be equal to $2.67 \times 10^{6}$ . \par

 Considering the signals events, which is equal to $\simeq 2.04 \times 10^4$ for $m_{H_2^{\pm}}= 1.3$ TeV and  $v_{\chi}=3.5$ TeV, then the statistical significance is $> 5.0 \sigma$, and for $m_{H_2^{\pm}}= 1.47$ TeV and same $v_{\chi}=3.5$ TeV the signals events is $\simeq 1.31 \times 10^3$, subsequently the statistical significance is $\simeq 0.80 \sigma$. Taking into account that for mass $m_{H_2^{\pm}}= 1.47$ and same $v_{\chi}=4.0$ TeV the signals events is $\simeq 8.35 \times 10^2$, it follows that the statistical significance is $\simeq 0.5 \sigma$. \par

\subsection{CLIC with 5 TeV }

For a charged Higgs mass greater than 1.5 TeV, it will be necessary to have a CLIC with energy greater than 3 TeV, so we will eventually do the phenomenology with a CLIC equal to 5 TeV \cite{ellis}. Let us suppose that the luminosity will be around $15$ ab$^{-1}$/yr, this is a conservative approach, that is, $3$ times greater than for the 3 TeV CLIC, considering that for the 1.5 TeV ILC, the 3 TeV CLIC is $13$ times greater. \par

Then the statistics we are expecting for this collider for $v_{\chi}=3.5(4.0, 5.0)$ TeV  and $m_{H_2^{\pm}}= 2.0$ TeV are $6.98 \times 10^5 (2.46 \times 10^5; 7.16 \times 10^4)$ of $H^{\pm}_{2}$ particles produced per year. In respect to the $m_{H_2^{\pm}}= 2.46$ TeV and the same parameters used above, it will give a total of $1.86 \times 10^4 (6.60 \times 10^3; 1.91 \times 10^3)$ of $H^{\pm}_{2}$, see Fig.8. The total widths for the vacuum expectation values given above are, $\Gamma_{H^{\pm}_{2}} = 3.09 \times 10^{-1}(2.56 \times 10^{-2}; 6.33 \times 10^{-4})$ for $m_{H_2^{\pm}}= 2.0$ TeV and $\Gamma_{H^{\pm}_{2}} = 2.34 (8.37 \times 10^{-1}; 1.74 \times 10^{-2})$ for $m_{H_2^{\pm}}= 2.46$ TeV, see Fig. 2. \par\par

Considering that the signal for $H^{\pm}_{2}$ are $E^{\pm} \nu(\bar{\nu})$ and taking into account that the BRs for these particles would be $BR(H^{\pm}_{2} \to E^{\pm} \nu(\bar{\nu} ) = 0.47 \% (7.90 \times 10^{-2}) \ \%) $, for the mass of the charged Higgs boson $m_{H^{\pm}}= 2.0(2.46)$ TeV and $v_{\chi}=3.5$ TeV, then  we would have approximately $\simeq 3.28 \times 10^{3} (15)$ events per year for CLIC with $5$ TeV, for the same signal as above, that is $E^{+} E^{-} \nu \bar{\nu}$, see Fig.6. \par

Considering that the second signal for $H^{\pm}_{2}$ are $Z V^{\pm}$ and taking into account that the BRs for these particles would be $BR(H^{\pm}_{2} \to Z V^{\pm} ) = 95.79 \ \% (95.81 \ \%) $, for the mass of the charged Higgs boson $m_{H^{\pm}}= 2.0(2.46)$ TeV and $v_{\chi}=3.5$ TeV, and that $V^{\pm}$ decay into $W^{\pm} Z$, whose $BR(V^{\pm} \to W^{\pm} Z ) = 99.20 \ \% (99.35 \ \%) $, see Fig. $9$ and \cite{morgan}, followed by leptonic decay of the boson $W^{\pm}$ into $\ell^{\pm} \nu(\bar{\nu})$, for which branching ratios for these particles would be $BR (W \to \ell^{\pm} \nu(\bar{\nu}))= 10.8 \%$ and $Z$ decay into $b \bar{b}$ whose branching ratios for these particles would be $BR(Z \to b \bar{b})= 15.2 \%$, then we would have approximately $1.66 \times 10^{3} (44 )$ events per year for
the signal $ b \bar{b} b \bar{b} b \bar{b} b \bar{b} \ell \ell \nu \bar{\nu}$. \par

Regarding $v_{\chi} = 4.0$ TeV, we have for the mass of the charged Higgs boson $m_{H_2^{\pm}}= 2.0(2.46)$ TeV, considering that the signal for $H^{\pm}_{2}$ are $E^{\pm} \nu(\bar{\nu})$ and taking into consideration that the BRs for these particles would be $BR(H^{\pm}_{2} \to E^{\pm} \nu(\bar{\nu} ) = 4.21 \% (1.66 \times 10^{-1} \%) $, then we would have roughly $1.04 \times 10^4 (11)$ events per year for the signal $E^{+} E^{-} \nu \bar{\nu}$, as shown in Fig.7. \par

Considering the signal for $H^{\pm}_{2}$ being $Z V^{\pm}$ and take into consideration that the BRs for these particles would be $BR(H^{\pm}_{2} \to Z V^{\pm} ) = 93.26 \ \% (95.81 \ \%) $, for the mass of the charged Higgs boson $m_{H_2^{\pm}}= 2.0(2.46)$ TeV, and that $V^{\pm}$ decay into $W^{\pm} Z$, whose $BR(V^{\pm} \to W^{\pm} Z ) = 99.36 \ \% (99.48 \ \%) $, see Fig. $9$ and \cite{morgan}, followed by leptonic decay of the boson $W^{\pm}$ into $\ell^{\pm} \nu(\bar{\nu})$, for which branching ratios for these particles would be
$BR (W \to \ell^{\pm} \nu(\bar{\nu}))= 10.8 \%$ and $Z$ decay into $b \bar{b}$ whose branching ratios for these particles would be $BR(Z \to b \bar{b})= 15.2 \%$, then we would have approximately $5.69 \times 10^{2} (16)$ events per year for the signal $ b \bar{b} b \bar{b} b \bar{b} b \bar{b} \ell \ell \nu \bar{\nu}$.   \par

With respect to vacuum expectation value $v_{\chi} = 5.0$ TeV, for the masses of $m_{H_2^{\pm}}= 2.0(2.46)$ TeV, Taking into account the same signal as above, that is, $E^{+} E^{-} \nu \bar{\nu}$ and considering that the branching ratios for $H^{\pm}_{2}$ would be $BR(H^{\pm}_{2} \to E^{\pm} \nu(\bar{\nu} ) = 100.00 \ \% (4.86 \ \%)$, then we would have $7.16 \times 10^4 (93)$ events per year for the signal $E^{+} E^{-} \nu \bar{\nu}$, as shown by Fig. 9. \par

Taking into consideration that the second signal for $H^{\pm}_{2}$ are $Z V^{\pm}$ and taking into account that the BRs for these particles would be $BR(H^{\pm}_{2} \to Z V^{\pm} ) = 0.0 \ \% (92.62 \ \%) $, for the mass of the charged Higgs boson $m_{H^{\pm}}= 2.0(2.46)$ TeV, and that $V^{\pm}$ decay into $W^{\pm} Z$, whose $BR(V^{\pm} \to W^{\pm} Z ) = 0.0 \ \% (99.57 \ \%) $, see Fig. $9$ and \cite{morgan}, followed by leptonic decay of the boson $W^{\pm}$ into $\ell^{\pm} \nu(\bar{\nu})$, for which branching ratios for these particles would be
$BR (W \to \ell^{\pm} \nu(\bar{\nu}))= 10.8 \%$ and $Z$ decay into $b \bar{b}$ whose branching ratios for these particles would be $BR(Z \to b \bar{b})= 15.2 \%$, then we would have approximately $0.0 (4)$ events per year for
the signal $ b \bar{b} b \bar{b} b \bar{b} b \bar{b} \ell \ell \nu \bar{\nu}$. We obtain the value of $BR(H^{\pm}_{2} \to Z V^{\pm} ) = 0.00$, because the $v_{\chi} = 5.0$ TeV restricts the mass of $m_{V^{\pm}}$  up to the value of $2380.0$ GeV.  \par

Taking into consideration the same backgrounds cited above and using the COMPHEP \cite{pukhov}, the total cross section for these processes for $5.0$ TeV will be equal to $2.31 \times 10^{-1}$ pb, consequently the number of backgrounds events will be equal to $3.47 \times 10^{6}$ . \par

Considering the signals events, that is $E^{+} E^{-} X$, which is equal to $\simeq 3.28 \times 10^{3} (15)$ for $m_{H_2^{\pm}}= 2.0 (2.46)$ TeV and $v_{\chi}=3.5$ TeV, then the statistical significance are $\simeq 1.76 (0.01) \sigma $, and for the same masses as above and $v_{\chi}=4.0$ TeV, the signals events are $1.04 \times 10^4 (11)$, then the statistical significance is $> 5.00(0.01) \sigma$. Considering the vacuum expectation value $v_{\chi}=5.0$ and masses $m_{H_2^{\pm}}= 2.0 (2.46)$ TeV the number of signals events are $\simeq 7.16 \times 10^4 (93)$, so the statistical significance are $> 5.00(0.05)\sigma$. \par

Taking into account the second signal for $H^{\pm}_{2}$, we would have approximately $1.66 \times 10^{3} (44 )$ events per year for the signal $ b \bar{b} b \bar{b} b \bar{b} b \bar{b} \ell \ell \nu \bar{\nu}$  for $m_{H_2^{\pm}}= 2.0 (2.46)$ TeV and $v_{\chi}=3.5$ TeV, then the statistical significance in this case are $\simeq 0.89 (0.02) \sigma $. For the same masses as above and $v_{\chi}=4.0$ TeV, the signals events are approximately $5.69 \times 10^{2} (16)$ events per year for the same signal as above, then the statistical significance are $0.31(0.01) \sigma$. For  $v_{\chi} = 5.0$ TeV and for masses of $m_{H_2^{\pm}}= 2.0(2.46)$ TeV, we have approximately $0.00 (29)$ events for these signals given above, then the statistical significance are $0,00(0.02)\sigma$. \par

In summary, we showed in this work that in the context of the $3-3-1$ model in some scenarios the signatures for charged Higgs boson $H^{\pm}$ can be significant in CLIC collider for both $3.0$ TeV and $5.0$ TeV. In other scenarios, the signal events will be small to be observed, therefore must be applied the cuts cited above and in other situations there are not signals events. It is worth mentioning that as the mass increases, in all cases the statistical significance decreases, this allows us to think that if the mass of the charged Higgs is large, we should think of more powerful accelerators. \par

\newpage

\begin{center}
FIGURE CAPTIONS
\end{center}


{\bf Figure 1}: Total cross section for the process $e^{-} e^{+} \rightarrow  H^{-}_{2} H^{+}_{2}$ as a function of $m_{H_{2}^{\pm}}$ at $\sqrt{s} = 3.0$ TeV for a) $v_{\chi} = 3.5$ TeV (solid line) and b) $v_{\chi} = 4.0$ TeV (point dashed line).


{\bf Figure 2}: The charged Higgs boson $H_{2}^{\pm}$ decay versus its mass for (a) $v_{\chi} = 3.5$ TeV (solid line), b) $v_{\chi} = 4.0$ TeV (point dashed line) and $v_{\chi} = 5.0$ TeV (short dashed line).


{\bf Figure 3}: The production cross section for the process $e^{-} e^{+} \rightarrow  H^{-}_{2} H^{+}_{2}$  for the resonance point $m_{Z^\prime} = 5976.43$ GeV for $v_{\chi}=3.5$ TeV. The solid line shows the cross section without the $ISR + BS$, and the
dashed line represents the $ISR + BS$ effect.

{\bf Figure 4}: The production cross section for the process $e^{-} e^{+} \rightarrow  H^{-}_{2} H^{+}_{2}$  for the resonance point $m_{Z^\prime} = 6830.21$ GeV for $v_{\chi}=4.0$ TeV. The solid line shows the cross section without the $ISR + BS$, and the
dashed line represents the $ISR + BS$ effect.

{\bf Figure 5}: The production cross section for the process $e^{-} e^{+} \rightarrow  H^{-}_{2} H^{+}_{2}$  for the resonance point $m_{Z^\prime} = 8539.47$ GeV for $v_{\chi}=5.0$ TeV. The solid line shows the cross section without the $ISR + BS$, and the
dashed line represents the $ISR + BS$ effect.

{\bf Figure 6}: Branching ratios for the charged Higgs decays as a function of $m_{H_{2}^{\pm}}$ for $v_{\chi}=3.5$ TeV.
 	
{\bf Figure 7}: Branching ratios for the charged Higgs decays as a function of $m_{H_{2}^{\pm}}$ for $v_{\chi}=4.0$ TeV.


{\bf Figure 8}: Total cross section for the process $e^{-} e^{+} \rightarrow  H^{-}_{2} H^{+}_{2}$ as a function of $m_{H_{2}^{\pm}}$ at $\sqrt{s} = 5.0$ TeV for a) $v_{\chi} = 3.5$ TeV (solid line), b) $v_{\chi} = 4.0$ TeV (point dashed line) and $v_{\chi} = 5.0$ TeV (short dashed line) .


{\bf Figure 9}: Branching ratios for the charged Higgs decays as a function of $m_{H_{2}^{\pm}}$ for $v_{\chi}=5.0$ TeV.


\begin{references}

\bibitem{wsg} S. Glashow, Nucl. Phys. {\bf 20} (1961) \ 579; A. Salam, in Elementary Particle Theory, ed. N. Svartholm, (1968); S. Weinberg, Phys. Rev. Lett. {\bf 19} (1967) 1264.

\bibitem{cms1} The CMS Collaboration, Phys. Lett. B $775$, 1 (2017)

\bibitem{supers} J. Wess and B. Zumino, Nucl. Phys. {\bf B70} (1974) 39.

\bibitem{cancell} J. Wess and B. Zumino, Phys. Lett. {\bf 49B} (1974) 52; J. Iliopoulos and B. Zumino, Nucl. Phys. {\bf B76} (1974) 310; S. Ferrara, J. Iliopoulos and B. Zumino, Nucl. Phys. {\bf B77} (1974) 413; E. Witten, Nucl. Phys. {\bf B188} (1981) 513.

\bibitem{sgut} S. Dimopoulos and H. Georgi, Nucl.Phys. {\bf B193} (1985) 150; S. Dimopoulos, S. Raby, and F. Wilczek, Phys. Rev. {D24} (1981) 1681; L. Iba\~nez and G. G. Ross, Phys. Lett. {\bf 105B} (1981) 439.

\bibitem{PT93a} V. Pleitez and M. D. Tonasse, Phys. Rev. D {\bf 48}, 2353  (1993).

\bibitem{PP92} F. Pisano and V. Pleitez, Phys. Rev. D {\bf 46}, 410 (1992);

\bibitem{FR92} P. H. Frampton, Phys. Rev. Lett. {\bf 69}, 2889 (1992).

\bibitem{DI05} A. G. Dias, Phys. Rev. D {\bf 71}, 015009 (2005).

\bibitem{glashow} S.L. Glashow, Nucl. Phys. B 22 (1961) 579; S. Weinberg, Phys. Rev. Lett. 19 (1967) 19;
A. Salam, in: Proceedings of the 8th Nobel Symposium, Editor N. Svartholm, Stockholm, 1968.

\bibitem{nilles} H. Nilles, Phys. Rept. 110 (1984) 1; H. Haber and G. Kane, Phys. Rept. 117 (1985) 75; R. Barbieri, Riv. Nuovo Cim. 11 (1988) 1.

\bibitem{TO96} M. D. Tonasse, Phys. Lett. B {\bf 381}, 191 (1996).

\bibitem{PT93} V. Pleitez and M. D. Tonasse, Phys. Rev. D {\bf 48}, 2353  (1993).

\bibitem{AK00} N. T. Anh, N. A. Ky and H. N. Long, {\it The Higgs sector in the minimal 3-3-1 model with the most general lepton-number conserving potential}, Report No. {\tt hep-ph/0011201}.

\bibitem{cieto02} J. E. Cieza Montalvo and M. D. Tonasse, Nucl. Phys. {\bf B623}, 325 (2002).

\bibitem{cieto03} J. E. Cieza Montalvo and M. D. Tonasse, Phys. Rev. D {\bf 67}, 075022 (2003).

\bibitem{ct2005} J. E. Cieza Montalvo and M. D. Tonasse, Phys. Rev. D {\bf 71}, 095015 (2005).

\bibitem{bhupal} P. S. Bhupal Dev, R. N. Mohapatra,  Phys. Rev. Lett. {\bf 115} (2015) 181803.

\bibitem{okada} arXiv:1811.11927v3 [hep-ph]

\bibitem{Nea10} J. Beringer {\it et al.} (Particle Data Group), Phys. Rev. D {\bf 86}, 010001 (2012).

\bibitem{tona96} M. D. Tonasse, Phys. Lett. B 381, 191 (1996); J.E.Cieza Montalvo, N.V. Cortez, and M. D. Tonasse, Phys. Rev. D 76, 117703 (2007).

\bibitem{atlas1} G. Aad {\it et al.}, Phys. Lett. B {\bf 710}, 49 (2012).

\bibitem{atlas11} G. Aad {\it et al.}(ATLAS Collaboration), Phys. Lett. B {\bf 716}, 1 (2012).

\bibitem{atlas2014} Phys. Rev. D {\bf 90}, 052005 (2014), arXiv:1308.5874v1 [hep-ex]; ATL-PHYS-PUB-2018-044.

\bibitem{kuraev} E. A. Kuraev and V. S. Fadin, Yad. Fiz. 41, 733 (1985)
[Sov. J. Nucl. Phys. 41, 466 (1985)]; O. Nicrosini and Luca Trentadue, Phys. Lett. B 196, 551 (1987); Pisin Chen, Phys. Rev. D 46, 1186 (1992); Kaoru Yokoya and Pisin Chen, Report No. KEK 91-2, 1991 (unpublished);
Orhan Cakir, New J. Phys. 8, 145 (2006).

\bibitem{morgan2} J. E. Cieza Montalvo {\it et al.}, (work in progress).

\bibitem{cheng} F. del Aguila, J. A. Aguilar-Saavedra, Nucl. Phys. B 813 (2009) 22; A. G. Akeroyd, Cheng-Wei Chiang, Naveen Gaur, JHEP 1011 (2010) 005.

\bibitem{pukhov} A. Pukhov et al., hep-ph/9908288.

\bibitem{ellis} arXiv:hep-ph/0412251v1 17 Dec 2004; See, for instance, accelconf.web.cern.ch, A $0.5$ TO $5$ TeV $e^{\pm}$ Compact Linear Collider, Jean-Pierre Delahaye {\it et al.} (unpublished).

\bibitem{morgan} Jorge. E. Cieza Montalvo {\it et al.}, MOMENTO, Revista de F\'isica 64, {\bf 16} (2022).









	
	
\end{references}
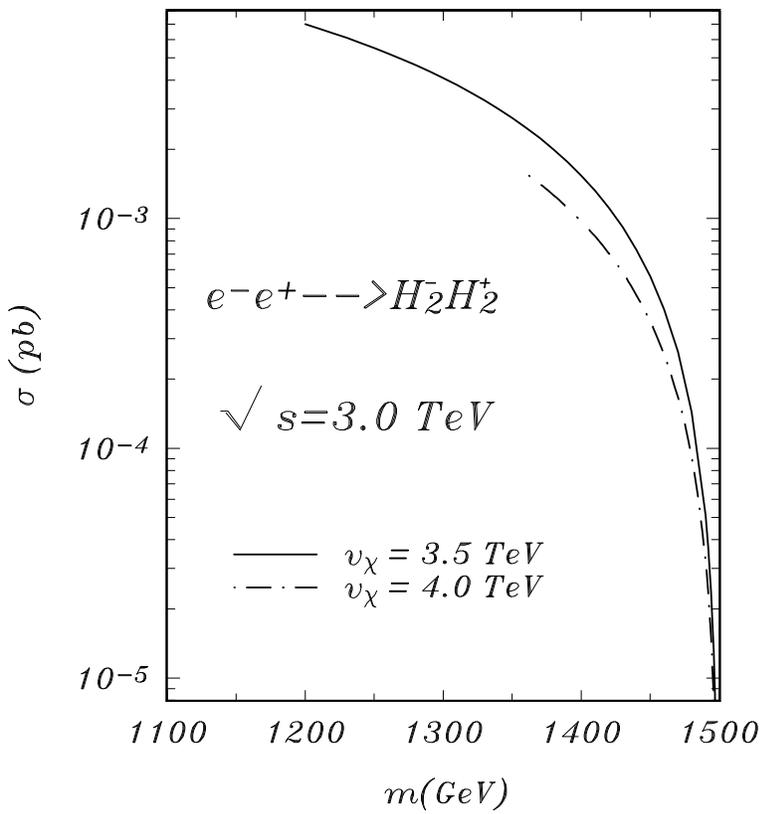
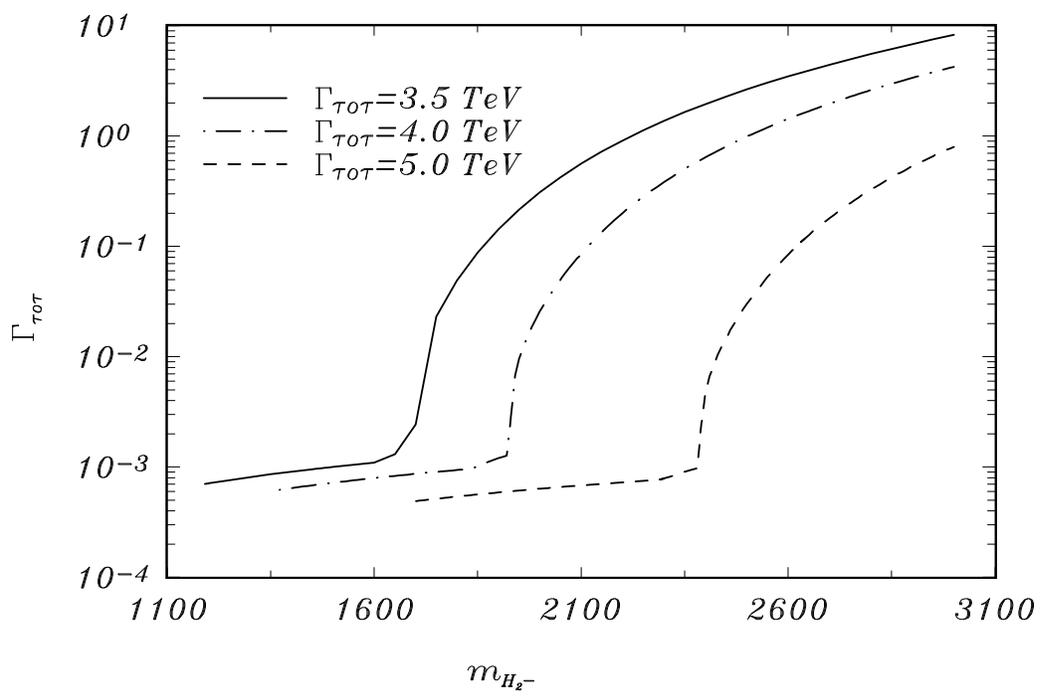
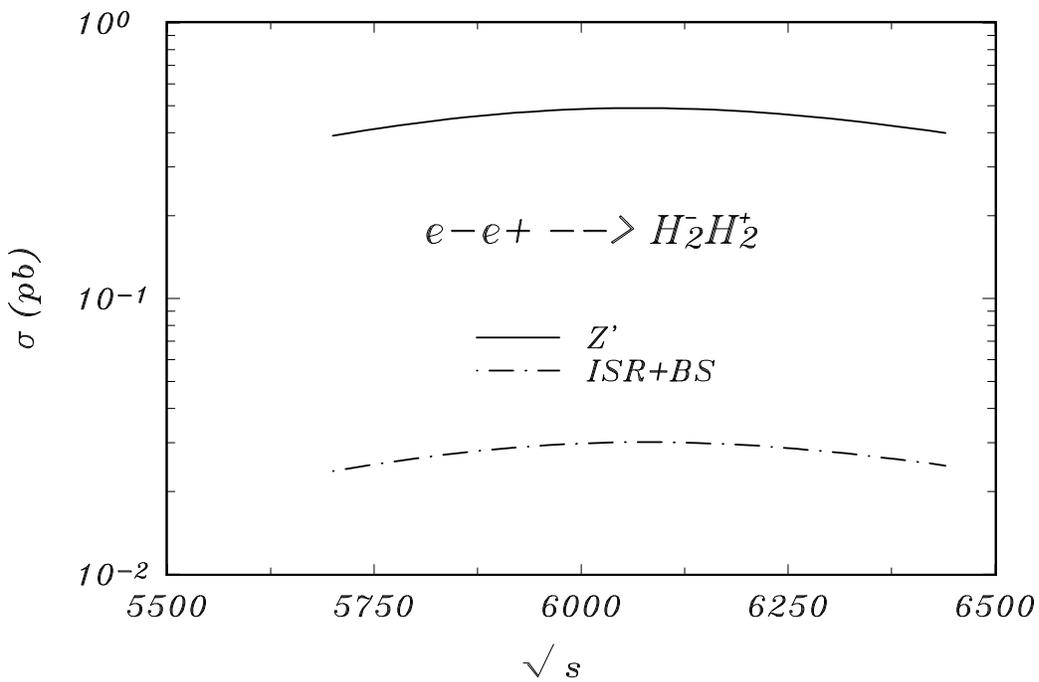
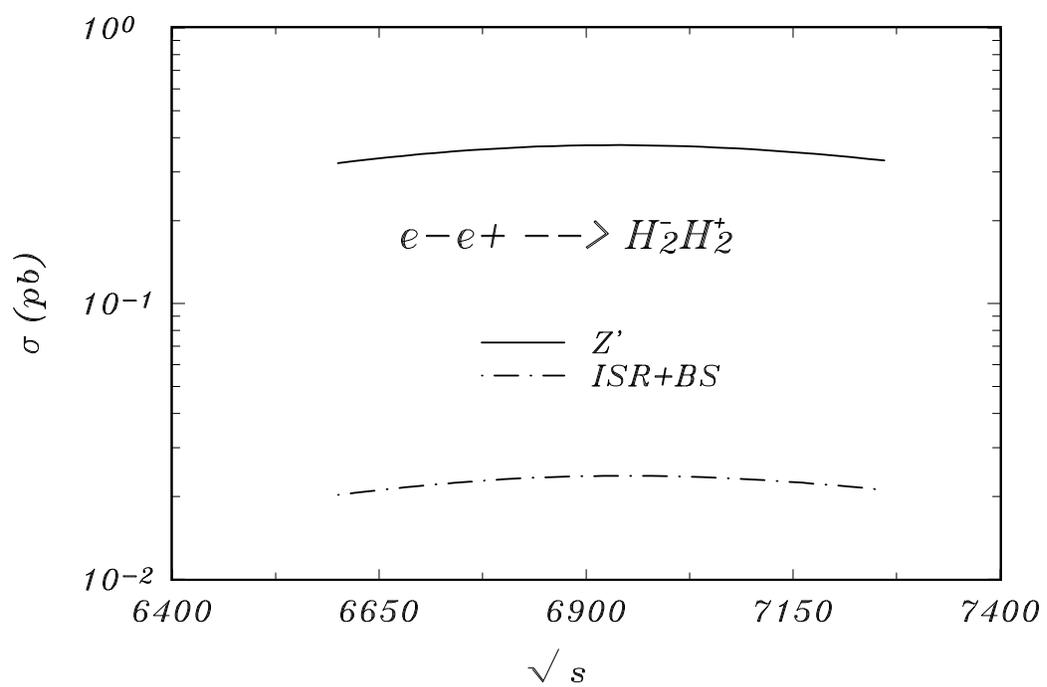
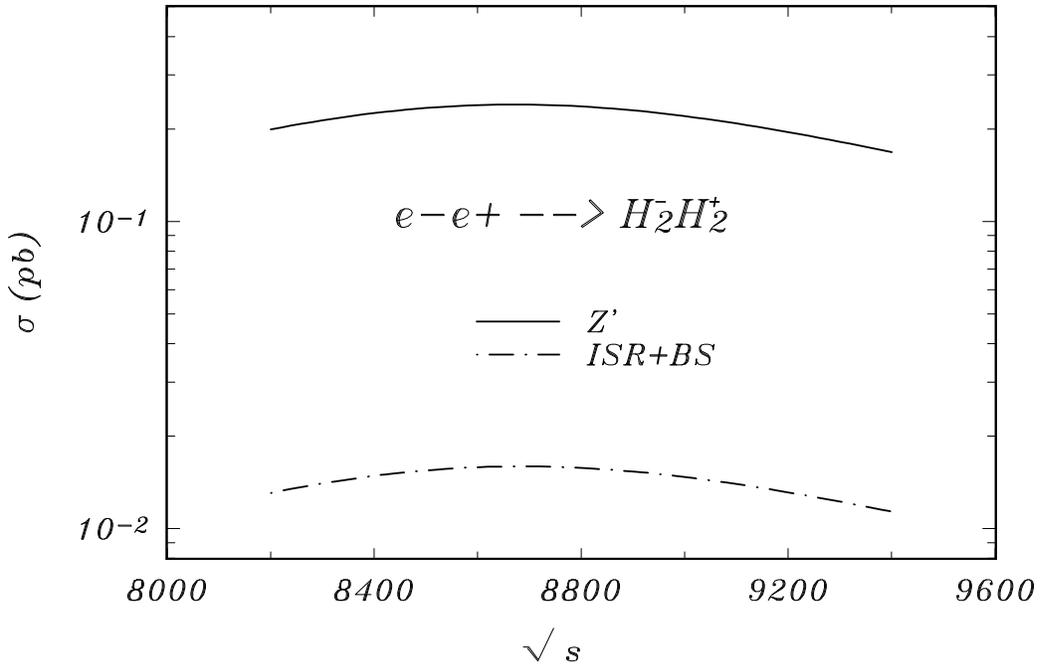
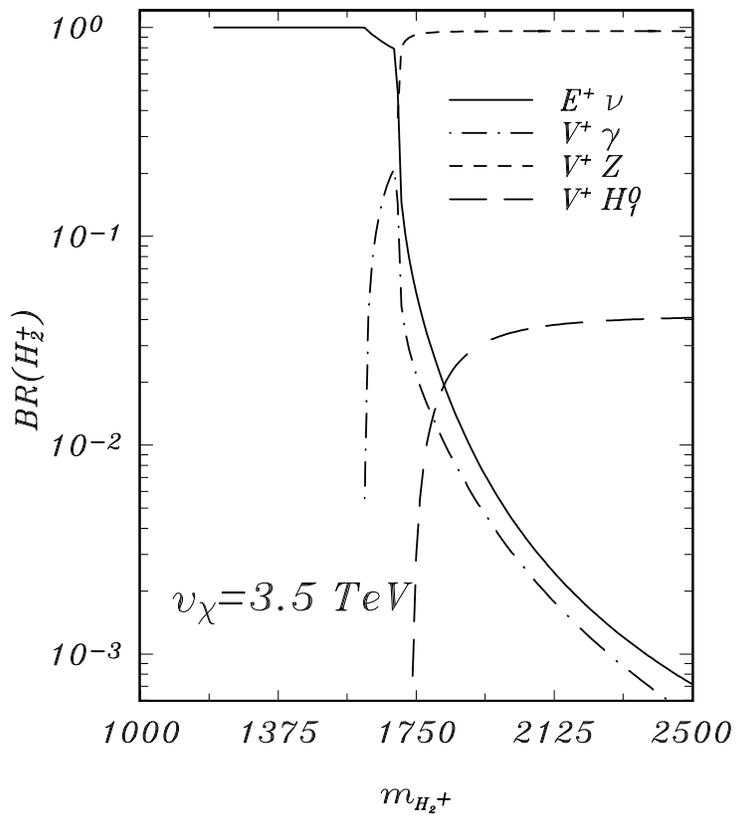
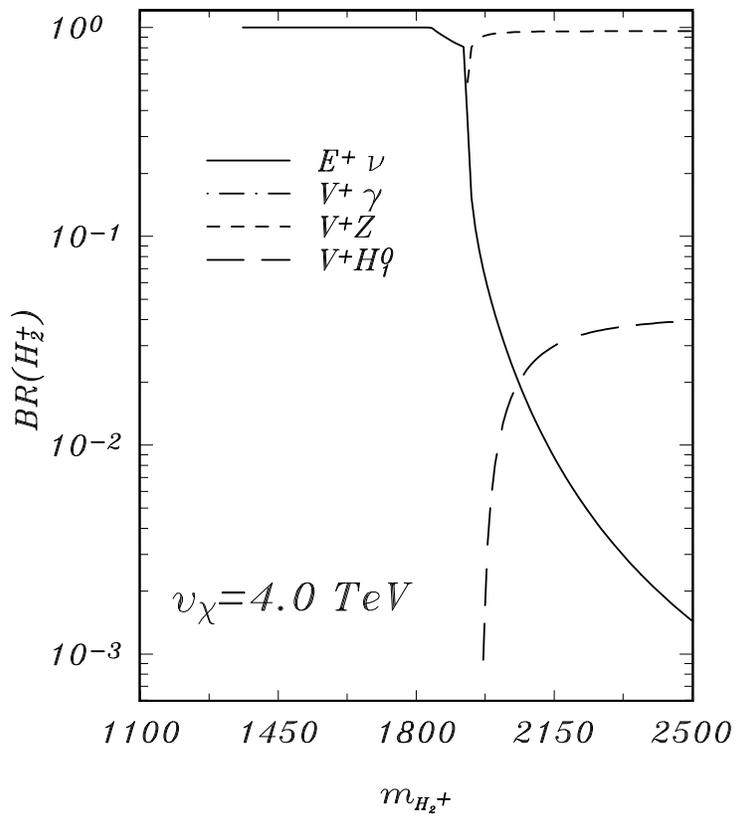
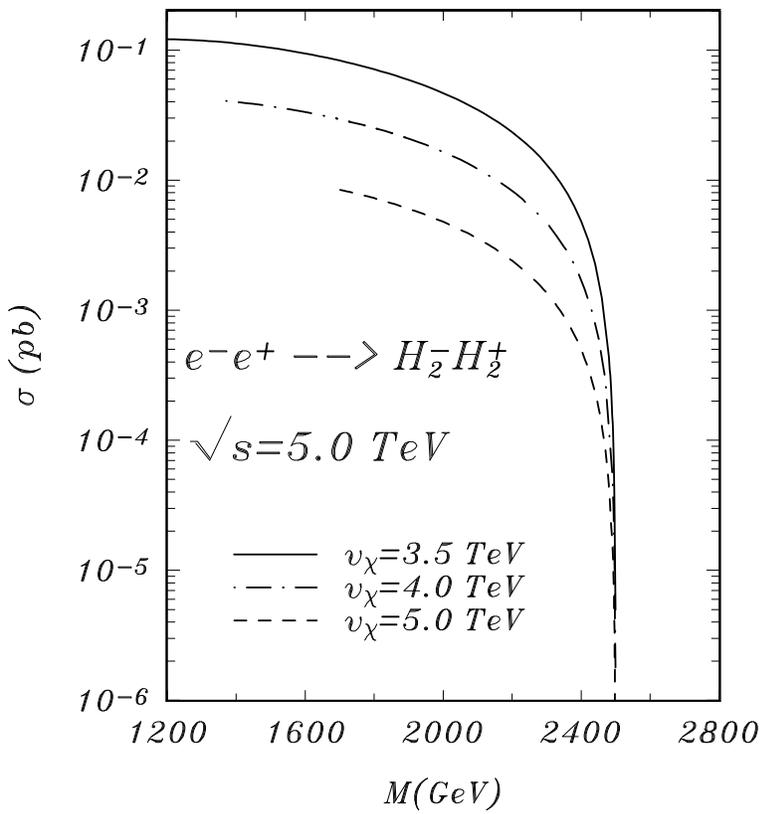
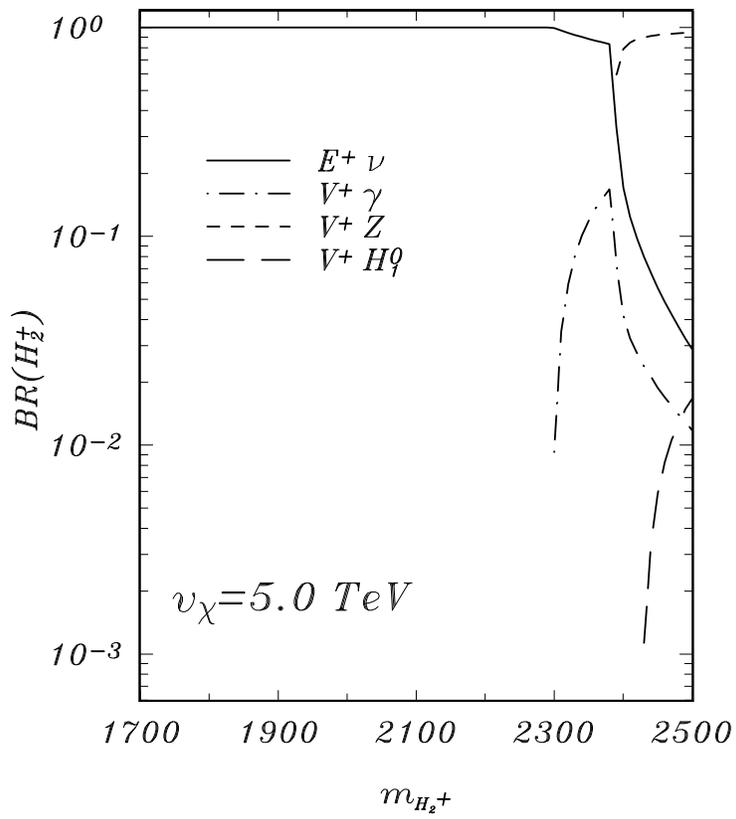
\end{document}